\documentclass[aps,prd,twocolumn,superscriptaddress,nofootinbib]{revtex4-1}

\usepackage{graphicx}
\usepackage{booktabs}
\usepackage[hidelinks]{hyperref}
\usepackage{color}
\hypersetup{
    colorlinks,
    linkcolor={blue},
    citecolor={blue},
    urlcolor={blue}
}
\usepackage{multirow} % Create tabular cells spanning multiple rows
\usepackage{amssymb,amsmath,amsfonts}
\usepackage[utf8]{inputenc}
\usepackage{nicefrac}
\usepackage{braket}
\usepackage{slashed}

\DeclareMathOperator{\tr}{tr}

\newcommand{\e}{\mathrm{e}}
\renewcommand{\d}{\mathrm{d}}

\def\Tc{T_{\mathrm{c}}}

\def\chitop{\chi_\mathrm{top}}
\def\Qlmax{Q^{\mathrm{L}}_{\mathrm{max}}}
\def\Qhmin{Q^{\mathrm{H}}_{\mathrm{min}}}
\def\Gmin{G_{\mathrm{min}}}
\def\Gmax{G_{\mathrm{max}}}
\def\Qph{Q'_{\mathrm{H}}}
\def\Qpl{Q'_{\mathrm{L}}}
\def\QMmin{Q^{\mathrm{M}}_{\mathrm{min}}}
\def\QMmax{Q^{\mathrm{M}}_{\mathrm{max}}}

\let\originalleft\left
\let\originalright\right
\renewcommand{\left}{\mathopen{}\mathclose\bgroup\originalleft}
\renewcommand{\right}{\aftergroup\egroup\originalright}

\begin{document}

\title{\boldmath
 Improved Reweighting for QCD Topology at High Temperature
}

\author{P. Thomas Jahn}
\email{tjahn@theorie.ikp.physik.tu-darmstadt.de}
\author{Guy D. Moore}
\email{guymoore@theorie.ikp.physik.tu-darmstadt.de}
\affiliation{Institut f\"ur Kernphysik (Theoriezentrum), Technische Universit\"at Darmstadt,\\ Schlossgartenstra{\ss}e 2, D-64289 Darmstadt, Germany}
\author{Daniel Robaina}
\email{daniel.robaina@mpq.mpg.de}
\affiliation{Max-Planck-Institut f\"ur Quantenoptik, Hans-Kopfermann-Stra{\ss}e 1, D-85748 Garching, Germany}

\begin{abstract}
In a previous paper \cite{Jahn:2018dke} we presented a methodology for
computing the topological susceptibility of QCD at temperatures where
it is small and standard methods fail.  Here we improve on this
methodology by removing two barriers to the reweighting method's
moving between topological sectors.  We present high-statistics,
continuum-extrapolated results for the susceptibility of pure-glue QCD
up to $7 T_c$.  We show that the susceptibility varies with
temperature as $T^{-6.7\pm 0.3}$ between $T=2.5\Tc$ and $T=7\Tc$, in
good agreement with expectations based on the dilute instanton gas
approximation.
\end{abstract}

\maketitle
\flushbottom

\section{Introduction \label{sec:Introduction}}

Two of the most interesting mysteries of particle physics are the
strong \emph{CP} problem and the origin of dark matter.  Both problems
could be solved simultaneously by the axion
\cite{Weinberg:1977ma,Wilczek:1977pj}. This is a light scalar particle
that is predicted by the Peccei-Quinn mechanism
\cite{Peccei:1977hh,Peccei:1977ur} and is a candidate for the dark
matter of the Universe. The additional degrees of freedom introduced
by the Peccei-Quinn mechanism also explain why the \emph{CP}
violating phase $\Theta_\mathrm{QCD}$ in the QCD Lagrangian vanishes.

The axion is therefore the subject of intense investigations both
experimentally and theoretically.  A theoretical prediction for the
axion's mass would be invaluable in the ongoing experimental search
for this particle (for a review on the experimental efforts we refer
to Ref.~\cite{Irastorza:2018dyq}).
If we assume that the axion makes up the dark matter and that
Peccei-Quinn symmetry was restored early in the Universe's history,
then this is possible \cite{Visinelli:2014twa}, but it requires
knowing the temperature history of the topological susceptibility of
QCD,
\begin{align}
\chi(T) = \int \d^4x \left\langle q(x)q(0) \right\rangle_T =
\frac{1}{V} \braket{Q^2},
\label{eq:chidef}
\end{align}
where $V$ is the Euclidean spacetime volume with periodic time
direction of extent $1/T$, and
\begin{align}
q(x) = \frac{1}{64\pi^2} \epsilon_{\mu\nu\rho\sigma} F_{\mu\nu}^a(x) F_{\rho\sigma}^a(x)
\end{align}
is the topological charge density and $Q = \int \d^4x \, q(x)$ the
topological charge.  In particular, the topological susceptibility
plays a nontrivial role in the axion abundance in the temperature
range from $3\: \Tc$ to $7 \: \Tc$ \cite{Moore:2017ond,Klaer:2017ond},
where $\Tc\simeq 155\,\mathrm{MeV}$ is the crossover temperature of
QCD.  Unfortunately calculations become very challenging
at high temperatures because topologically nontrivial configurations
are very suppressed.  In a lattice study in finite volume, only a tiny
fraction of configurations will possess topology, which makes it
challenging to achieve good statistics in a conventional Monte-Carlo
study.  In the last years there was
a lot of progress in studying topology at high temperatures
\cite{Frison:2016vuc,Berkowitz:2015aua,Taniguchi:2016tjc,%
  Bonati:2015vqz,Petreczky:2016vrs,Borsanyi:2015cka,Bonati:2018blm}. In
particular, Borsanyi {\textit{et al}} have proposed one way around the
difficulty in sampling topology at high temperatures, by performing
separate simulations in the instanton-number 0 and instanton-number 1
ensembles over a range of temperatures \cite{Borsanyi:2016ksw}.
In Ref.~\cite{Jahn:2018dke} we presented an alternative, more direct
method that allows us to study topology up to high temperatures, by
using a reweighting technique.

Our overall strategy will be the same as in \cite{Jahn:2018dke}.  We
consider temperatures such that topology is rare; almost all
configurations have $Q=0$, a small fraction have $Q=\pm 1$, and
$|Q| \geq 2$ is so suppressed that it plays a negligible role.  We
assign topology based on whether the bosonic determination of $Q$,
measured after a certain depth of gradient flow
\cite{Narayanan:2006rf,Luscher:2009eq}, exceeds a threshold
value.  The susceptibility is
$\chi(T) = \langle Q^2 \rangle/a^4 N_\tau N_x N_y N_z$, with $a$ the
lattice spacing,
$N_\tau$ the number of lattice points in the temporal direction, and
$N_{x,y,z}$ the number of points across each spatial direction.

The core idea of our approach is to overcome the small fraction of
configurations which have topology, by sampling the configuration
space according to the modified distribution
\begin{align}
\d P_\mathrm{rew}(U) = \frac{\e^{- \beta S_\mathrm{W}[U]+W(\xi)}\mathcal D U}
   {\int \mathcal DU \e^{-\beta S_\mathrm W[U] + W(\xi)}}
\end{align}
instead of using the standard distribution
$e^{-\beta S_{\mathrm{W}}[U]} \mathcal{D} U$. By choosing the
\emph{reweighting function}
$W(\xi)$ and the \emph{reweighting variables} $\xi$ appropriately,
topologically non-trivial
configurations can be artificially enhanced in the sample, and the
barriers between topological sectors, which lead to large
autocorrelations in the topology, can also be overcome.  To account
for the modified weight, the result is computed using the following
definition of the expectation value:
\begin{align}
  \left\langle O\right\rangle = \frac{\sum^N_i O_i\e^{-W(\xi_i)}}
               {\sum^N_i \e^{-W(\xi_i)}} \,.
\end{align}
Since our reweighting variables $\xi$ are rather nontrivial, the
inclusion of $e^{W[\xi]}$ in the sampling weight is achieved by a
Metropolis accept-reject step.  In \cite{Jahn:2018dke}
we presented an automated way to build an optimized choice for the
reweighting function.  We also argued for the use, as reweighting
variable $\xi$, of the absolute value of the bosonic
definition of topological number $Q$, measured using an $a^2$-improved
definition of $q(x)$ and after a modest amount of gradient flow.  This
choice is \textsl{not} truly topological, and to emphasize that fact,
we will write it as $Q'$ rather than as $Q$.  The choice of a
non-topological measurable is
deliberate; $Q'$ takes values near $Q'=0$ for regular non-topological
configurations, values near $Q'=1$ for configurations with instanton
number $\pm 1$, and values around $Q' \sim 0.5$ for ``dislocations,''
small knots of $q(x)$ which are the intermediate states, on the
lattice, between topological and nontopological gauge configurations.
We found that gradient flow depths $t_f \simeq 0.42a^2$ seem to work
well.

\begin{figure}[t]
\centering
\includegraphics[width=\linewidth]{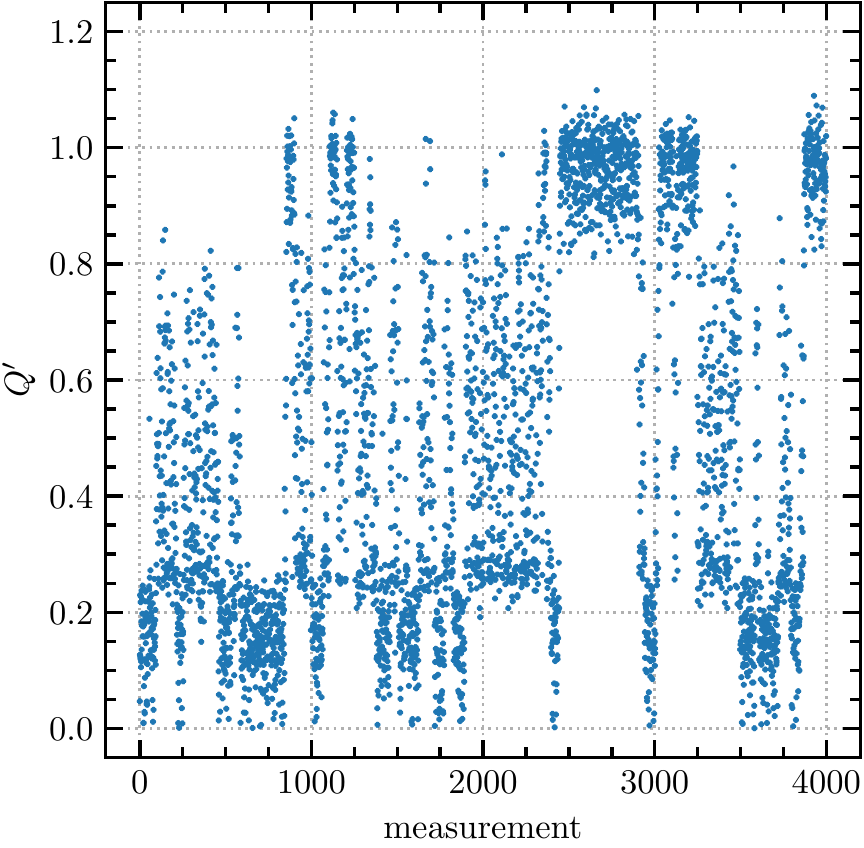}
\caption{Piece of a Markov-chain history of $Q'$ against measurement
  number for lattice B$_{\mathrm{2d}}$. The reweighting allows
  efficient sampling in three regions, $0<Q'<0.27$, $0.20<Q'<0.85$,
  and $0.8<Q'<1.05$, but has difficulty moving between these
  regions. The same figure is published in our paper
  \cite{Jahn:2018dke}.}
\label{fig:chainbit}
\end{figure}

Using the reweighting approach as developed and established in
Ref.~\cite{Jahn:2018dke}, we observed that there are three regions
where reweighting allows for efficient sampling, while it has problems
moving between those regions, as indicated in
Fig.~\ref{fig:chainbit}. The problem of those two ``barriers'' gets
severe when we go to higher temperatures or finer lattices. Solving
those problems would therefore give significantly better efficiency;
the susceptibility could be determined from much shorter Markov
chains.  This will be important in the future, when we move from our
current exploratory pure-glue, Wilson-action studies to studies
including light fermions.  The reason for the
occurrence of the barriers is that reweighting only in terms of the
topological charge is incomplete and missing some information that
distinguishes between the different regions. In this paper, we address
how to overcome both barriers. This leads to improved efficiency of
the reweighting method and allows for a direct measurement of the
topological susceptibility up to very high temperatures and fine
lattices.

This paper is organized as follows: In Sec.~\ref{sec:globmethod}, we
discuss in detail the modification of the original reweighting
approach that significantly improves the efficiency of the
method. Sec.~\ref{sec:globresults} contains the lattice determination
of the susceptibility at three temperatures,
$T = 2.5 \,\Tc$, $4.1\,\Tc$, and $7.0\,\Tc$, each at three spacings
with $N_\tau = 10,12,14$.  This allows us to check our previous
results and to perform a continuum extrapolation and a power-law fit
as a function of temperature.
A discussion of our results can then be found in
Sec.~\ref{sec:globdiscussion}.

\section{The Method \label{sec:globmethod}}

In this section, we discuss the improvement of the reweighting method
that overcomes both ``barriers.''  We shall refer to those barriers as
the \emph{low barrier}, \textit{i.e.}, the barrier at around $Q'
\simeq 0.25$ in Fig.~\ref{fig:chainbit}, and the \emph{high barrier},
\textit{i.e.}, the barrier at around $Q' \simeq 0.85$ in
Fig.~\ref{fig:chainbit}. This section starts by addressing the origin
of both problems and how to improve tunneling through the
corresponding barriers. We shall find that both problems need to be
solved differently and we hence have to split up the whole lattice
setup into multiple distinct Monte Carlo samples. This section is
concluded by a discussion of how the different Monte Carlo samples can
be combined to a measurement of the topological susceptibility.

\subsection{The Low Barrier}
The low barrier occurs because the algorithm has problems to move
between configurations with trivial topology and dislocations,
\textit{i.e.}, small concentrations of topological charge that are the
intermediate steps between the $Q=0$ and $Q=1$ sectors. For an
additional reweighting, we therefore need a quantity that
distinguishes between these two types of configurations. Since for
dislocations the
topological charge is spatially very concentrated, we expect that the
action is also sharply peaked at the dislocation's location.  We
therefore consider the peak action density
\begin{align}
G \equiv \max_{\tilde x} \big\{ S(\tilde x) \big\} \,,
\end{align}
where $\tilde x$ denotes a point in the dual lattice and
\begin{align}
S(\tilde x) = \sum_{P(\tilde x)} \mathrm{Re}\tr \left( 1 - P(\tilde x) \right)
\end{align}
with $P(\tilde x)$ being the 24 plaquettes that lie on the hypercube
bounding the primitive cell with center $\tilde x$.
We use $G$ as a quantity that distinguishes between topologically
trivial configurations and dislocations.%
\footnote{%
  We name the peak
  action density $G$ for ``globbiness'' because this quantity
  determines how ``globby'' in the sense of spatially concentrated a
  configuration is.}
Note that, by definition,
\begin{align}
\sum_{\tilde x} S(\tilde x) = 4 \cdot S_\mathrm W \,,
\end{align}
where $S_\mathrm W$ is the Wilson gauge action.

\begin{figure}
  \centering
  \includegraphics[width=0.46\linewidth]{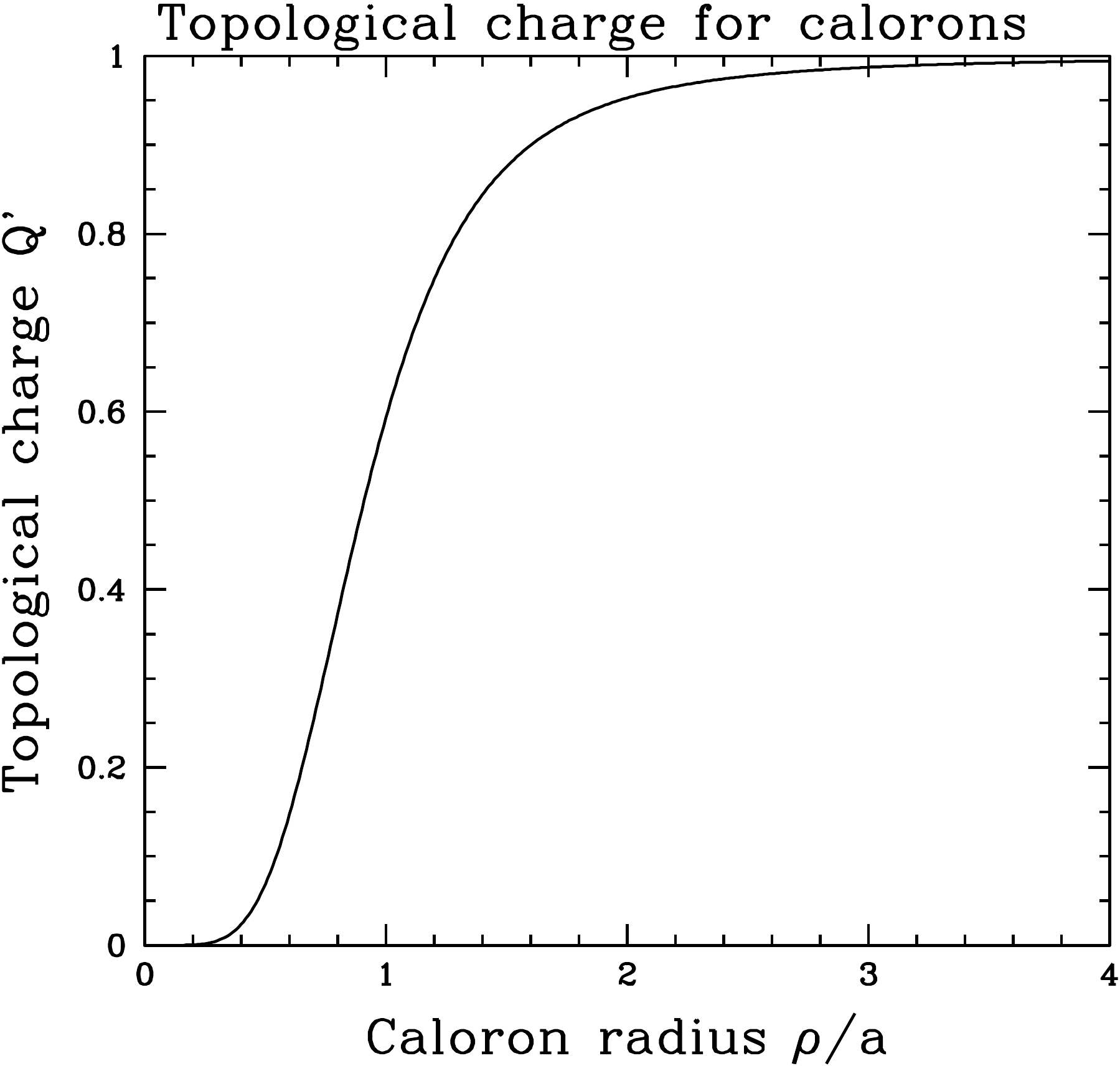}
  \hfill
  \includegraphics[width=0.49\linewidth]{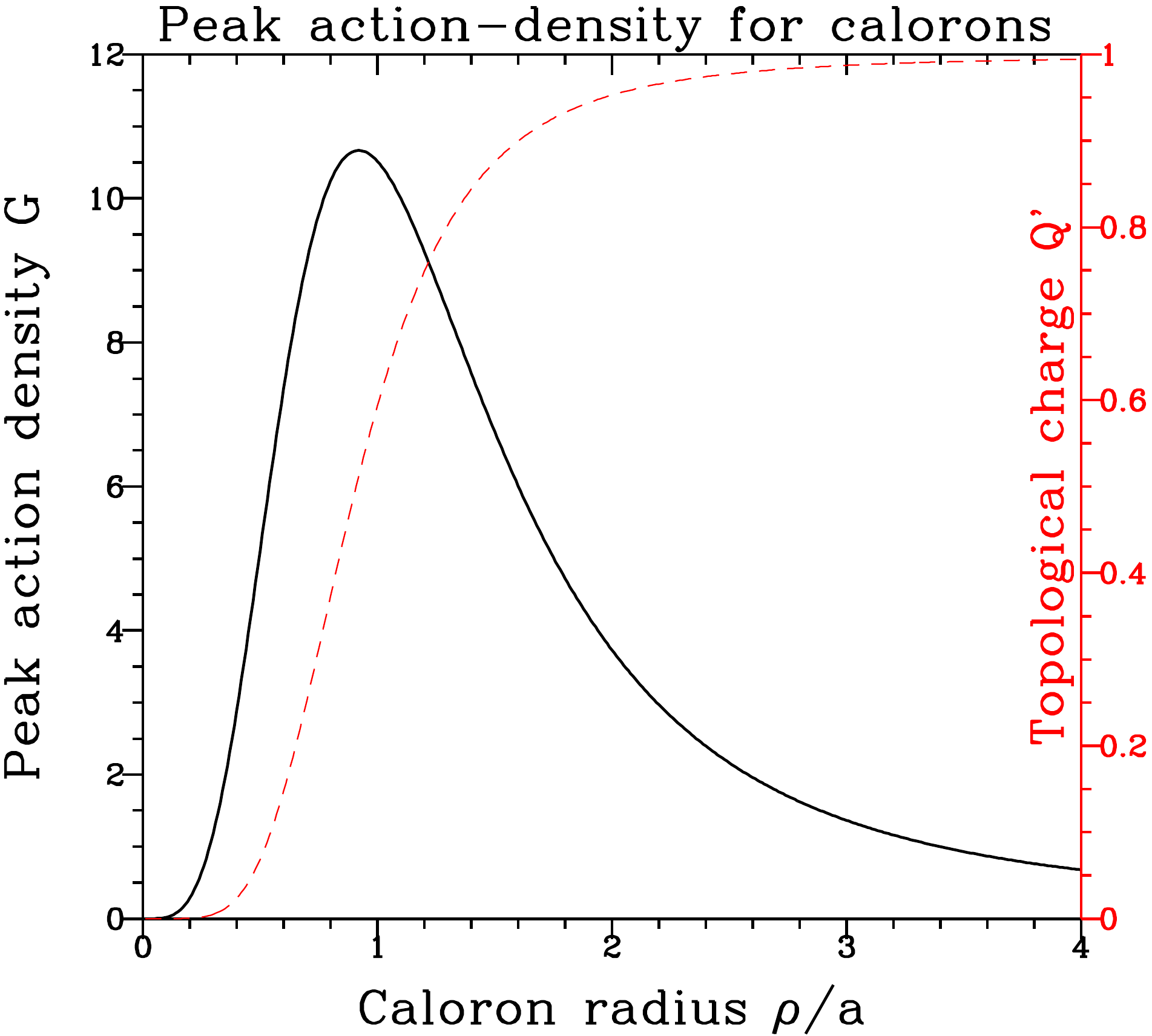}
\caption{
  Left: Symanzik-improved topological charge $Q$ for
  Harrington-Shepard (HS) calorons as a function of caloron size.
  Right:  peak action density $G$ for the same HS calorons (with
  $Q$ also shown in red).
  The maximum in $G$ occurs precisely where $Q$ is intermediate
  between 0, its $\rho=0$ value, and 1, its large-$\rho$ asymptote.}
\label{fig:globbiness}
\end{figure}

Figure \ref{fig:globbiness} shows the values of the topological charge
$Q$ and this action-density measure $G$ for lattice-discretized
Harrington-Shepard (HS) calorons, as a function of the caloron
radius $\rho/a$ (cf.\ Ref.~\cite{Jahn:2018jvx}). As
expected, $G$ peaks at $\rho \simeq a$,
\textit{i.e.}, for dislocations, precisely where $Q'$ is intermediate
between topology-0 and topology-1 values.  On the other hand, for
configurations where the caloron is so small that it falls between the
lattice points, $G$ is also small; and $G$ is also small for large
calorons which clearly display topological character.  Therefore $G$
is a good discriminant for configurations for which topology is
ambiguous.

\begin{figure}
  \centering
  \includegraphics[width=0.58\linewidth]{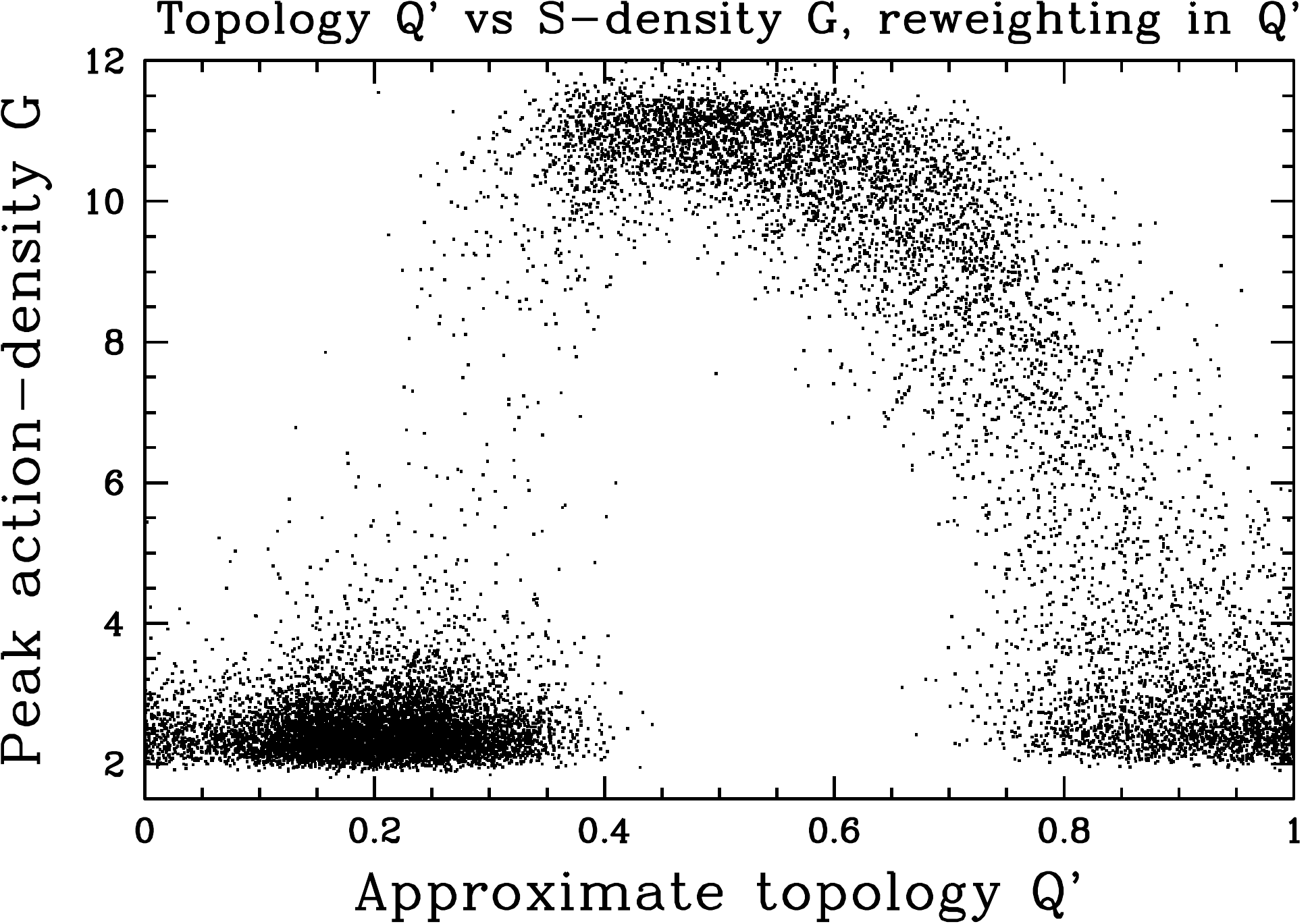}
  \hfill
  \includegraphics[width=0.38\linewidth]{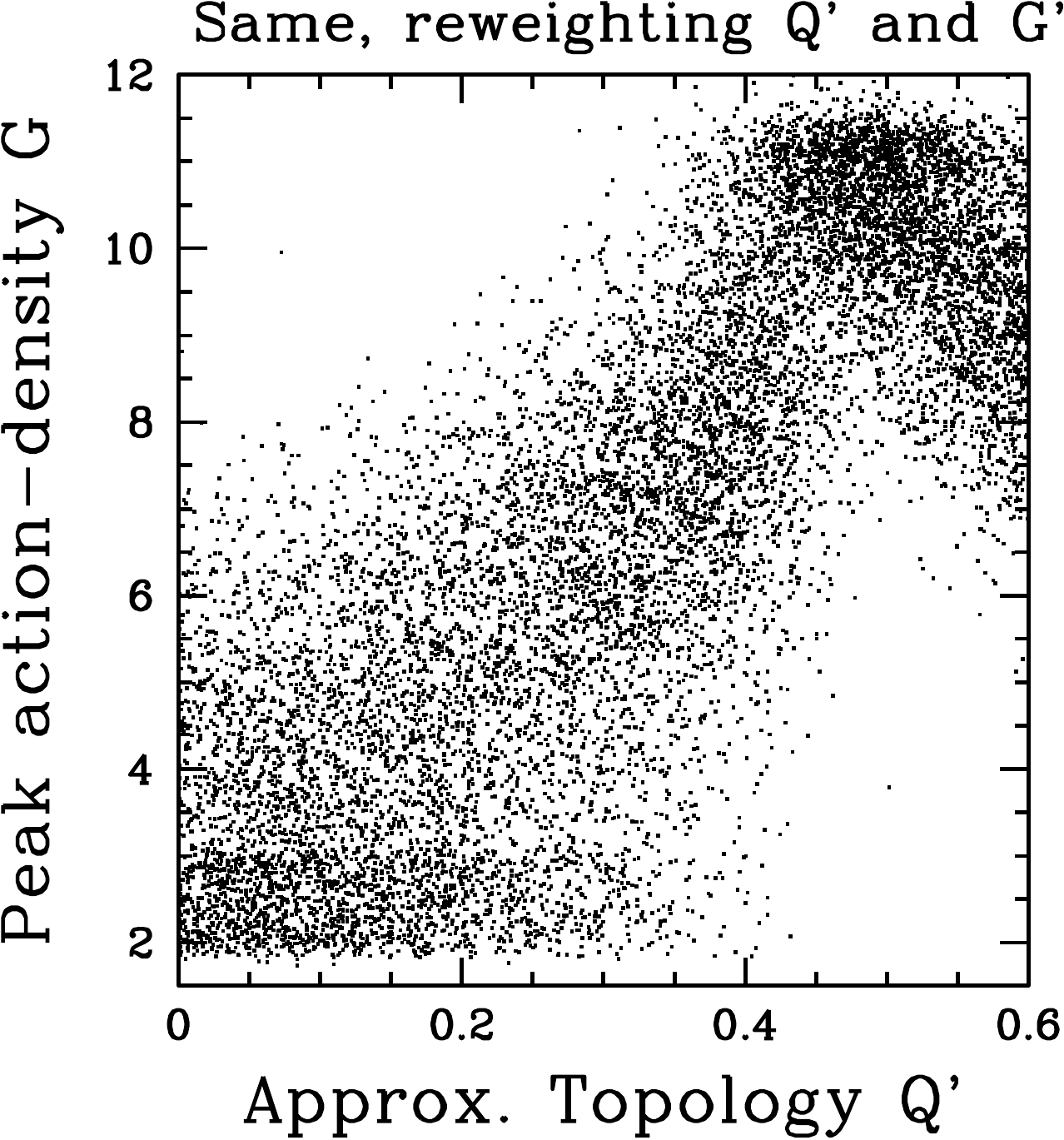}
  \caption{
    Distribution of topological charge $Q'$ and peak action density
    $G$, obtained during a reweighted Markov chain.
    Left: reweighting in terms of $Q'$ only (on an $8\times 32^3$
    lattice at $T=4.1\Tc$).  Right:  reweighting in terms of both $Q'$
    and $G$ (on lattice $B_1$ and only extending up to $Q'=0.6$).
    \label{fig:Gbeforeafter}
  }
\end{figure}

The reason for the ``low barrier'' between $Q'<0.2$ configurations
and $Q'>0.3$ configurations, seen in Figure \ref{fig:chainbit}, is
precisely because of the difficulty in getting between non-topological
configurations and ambiguous-topology configurations characterized by
a large $G$ value.  To see this, consider the distribution, in the
$(Q',G)$ plane, of the configurations generated by a $Q'$-reweighted
HMC Markov chain, shown in the left panel of Figure
\ref{fig:Gbeforeafter}.
There is a clear ``gap'' in the sample, with very few points sampling
the region around $G=7$ and $Q'=0.25$.  For every $Q'$ value between 0
and 0.5, the sample is dominated either by%
\footnote{%
  $G$ is always nonzero because of the action associated with ordinary
  fluctuations, which pervade the lattice.  This ``background'' level
  of $G$ is dependent on $\beta$, the lattice volume, and especially
  on the gradient flow depth $t_f$.  One criterion for $t_f$ is that
  it be sufficient that this background $G$ value is far below
  $G\simeq 11$, the value for a dislocation.}
$G\simeq 2$ configurations or by $G \sim 11$ configurations, or by
a linear combination of these two; it is never controlled by
configurations intermediate between these $G$ values.  The failure to
sample such configurations inhibits the Markov chain's ability to
sample both parts of the configuration space.

To encourage transitions across this ``gap,'' instead
of reweighting solely in terms of $Q'$ (here renamed $\Qpl$ for
reasons which will become clear), we perform an additional
reweighting in terms of $G$. The effective reweighting function is
then the sum of two individual reweighting functions:
\begin{align}
W_\mathrm{low}(\Qpl,G) = W_Q(\Qpl) + W_G(G) \,.
\end{align}
Note that $G$ is also evaluated after some amount of gradient flow
$t_{f,G}$ that in principle does not have to equal the flow time
$t_{f,\mathrm L}$ after which $\Qpl$ is evaluated. Hand-tuning
showed that the amount of gradient flow that gives the best
performance depends on the size of the lattice; larger lattices need
more flow. The specific choices of these parameters for our lattices
are listed in Tab.~\ref{tab:ourlattsglob}. We also saw that HMC
trajectories with one step of length $0.2 \, a$ give the best
performance in this region%
\footnote{%
  Dislocations are very small objects which are sensitive to
  relatively small changes in the gauge-field links.  The reweighting
  in terms of $Q'$ and $G$ are implemented via an accept-reject step,
  and too-large changes to the fields lead to a high reject rate.
  Therefore in this region the HMC trajectories have to be very
  short.}.
Both reweighting functions are built simultaneously in the same way as
described in detail in Ref.~\cite{Jahn:2018dke}.

Reweighting in terms of both $\Qpl$ and $G$ removes the barrier,
as seen in the right panel of Figure \ref{fig:Gbeforeafter}.  It
increases by more than a factor of 5 the number of transitions between
$\Qpl \simeq 0$ and $\Qpl \simeq 0.5$ configurations, for a given
number of HMC trajectories.  Therefore this approach appears to cure
the low barrier.

Unfortunately reweighting in terms of $G$ does not significantly help
with the high barrier, as already suggested in the left panel of
Figure \ref{fig:Gbeforeafter}.  Therefore we will only use this method to
perform a Monte-Carlo over a reduced range of $\Qpl$ values, which we
will call the \emph{low region} (L).  Specifically, we reweight $\Qpl$
only up to a value $\Qlmax$, which we choose to be 1.15 times the $\Qpl$
value for the HS caloron with the maximum $G$ value, which we can look
up from the right panel of Figure \ref{fig:globbiness}.  We strictly
reject configurations with $\Qpl > \Qlmax$; the role of such
configurations will be covered by the middle and upper regions, which
we will describe next.
All values of $G$ are allowed, but our reweighting function is only
nontrivial between $\Gmin$ and $\Gmax$, where $\Gmin$ is the mean
value we find in a short non-reweighted Markov chain and $\Gmax$ is
the largest value for the caloron solutions shown in the left panel of
Figure \ref{fig:globbiness}.  Note that configurations that are
outside of this $G$-interval are not rejected; we just extend the
reweighting function as a constant beyond the limits, that is,
$W_G(G>\Gmax) = W_G(\Gmax)$ and $W_G(G<\Gmin) = W_G(\Gmin)$.

\subsection{The High Barrier}

The actual configurations carrying topology at finite temperature are
expected to be nontrivial objects with large fluctuations, not ``clean'' HS
calorons.  Nevertheless, we will refer to them as calorons in what
follows.  We can define the ``size'' of such a configuration as the
size of the HS caloron it approaches under gradient flow -- note that,
up to lattice artifacts, HS calorons are extrema of the action and do
not change size under gradient flow, so gradient flow brings
topological objects towards clean HS calorons.
Therefore it makes logical sense
to discuss the size distribution of calorons.  Perturbatively we
expect the dominant size to be $\rho \sim 0.5 N_\tau a$
\cite{Jahn:2018jvx}; but the size of a dislocation is closer
to $\rho \sim 1.5 a$.  Since we want to compare the relative weight of
topological and non-topological configurations, and $\rho \sim 1.5a$
configurations are a necessary intermediate step, we have to make sure
that our HMC Markov chain moves efficiently across the range of
caloron sizes from $\rho \sim 1.5 a$ to $\rho \sim 0.5 N_\tau a$.
Our continuum extrapolation will involve $N_\tau = 10,12,14$, so we
need efficiency up to quite large calorons (in lattice units).
As we understand it, the difficulty in doing so is what drives the
high barrier in Figure \ref{fig:chainbit}.

The quantity $G$ that we introduced for the low barrier is
unfortunately not helpful here, because there is no real
``gap'' in the $(Q',G)$ distribution between $Q'=0.5$ and $Q'=1$ in
the left panel of Fig.~\ref{fig:Gbeforeafter}. Until now we were not
able to find an auxiliary variable which significantly improves
performance in this high region.  Therefore we will have to find other
ways to make this region more efficient.

The first thing to note is that the calorons under consideration
are rather large and robust objects.  Therefore it is no
longer necessary to perform the Markov chain using short, inefficient
HMC trajectories.  So we switch to longer HMC trajectories (10 steps
of $0.25 \, a$), which induce much larger changes in our gauge field
configuration.  This already significantly improves efficiency in
the high region.

Next, consider the variable $Q'$ we use for reweighting.  The right
panel of Figure \ref{fig:globbiness} shows how the $a^2$-improved
definition of topology
varies as a function of caloron size for clean, idealized HS
calorons.  We see that in the range of interest, $\rho \in [1.5,7]a$,
$Q$ varies very little.  Therefore, we need a precise determination of
$Q$ if it is to prove useful in distinguishing between different
caloron sizes.  Unfortunately, $Q'$ is a noisy measurable.
To see why a little better, note that our lattice definition
of $Q'$ is contaminated by high-dimension operators:
$q_{\mathrm{latt}}'(x) \simeq q(x) + c_2 a^2 F^a_{\mu\nu} D^2 \tilde F^a_{\mu\nu}
+ c_4 a^4 F^a_{\mu\nu} D^4 \tilde F^a_{\mu\nu} + \ldots$.
Operator improvement forces $c_2=0$, but higher terms still exist,
and the dimension-8 operators appearing in this expansion do not
integrate to topological invariants.  Contributions from these high
dimension operators are dominated by the shortest-distance scale which
is not erased by gradient flow.  Therefore, integrating up
$q_{\mathrm{latt}}'(x)$ will give the topology of the configuration,
plus nontopological fluctuations which are suppressed by
$\sim a^4/t_f^2$, but whose variance is extensive in the lattice
volume.  This makes it clear that a larger depth of flow can greatly
suppress these fluctuations, providing a cleaner value of $Q'$.
So defining $Q'$ using a
larger amount of gradient flow leads to a cleaner variable, which
is better able to distinguish between different caloron sizes.
Unfortunately, a larger amount of gradient flow simply destroys
dislocations, so this approach cannot be used in the low region.
Therefore we will use one amount of gradient flow
$t_{f,\mathrm{low}} \in [0.36,0.54]a^2$
to define $\Qpl$, which we will use
in the low region, and another amount of gradient flow
$t_{f,\mathrm{high}} \in [0.96,1.32]a^2$ to define $\Qph$,
which we will use in the high region.  Larger lattice volumes and
lower temperatures demand larger $t_f$ values; for the smallest
lattices we consider, we choose the lower ends of the indicated ranges,
while for the largest lattices we use the upper ends.

Finally, when we reweight in terms of some variable, our methodology
involves choosing \textsl{intervals} in that variable, with $W[\Qph]$
chosen as piecewise linear across each interval.  Our method for
determining the reweighting function $W[\Qph]$ leads to approximately
equal sampling of each interval.  We want to sample nearly equally in
$\rho$, not in $\Qph$.  Therefore, we choose a uniform set of $\rho$
values, from a minimum value for which
$\Qph=0.7 \equiv \Qhmin$, up to a maximum value
$\rho_{\mathrm{max}} = 0.5 N_\tau a$.  We use the right-hand plot in Figure%
\footnote{%
  Technically we have to re-make the plot, applying
  $t_{f,\mathrm{high}}$ depth of gradient flow before measuring
  $\Qph$.  This modifies the small-$\rho$ part of the plot but has
  almost no influence at larger $\rho$, see \cite{Jahn:2018jvx}.}
\ref{fig:globbiness} to look up the associated $\Qph$ value of each, and use
these as the edges of our intervals.  This leads to more sampling, and
more sensitivity, at the largest $\Qph$ values, corresponding to larger
calorons.

The price we pay for these methodological changes is, that our
procedure now differs -- both in HMC trajectory choice and in
reweighting variable choice -- between the upper and lower regions.
And as we have defined them, these regions do not necessarily even
overlap.  Therefore we will need a procedure for ``sewing together''
these regions, which we describe next.

\subsection{The Middle Region}
\label{sec:middleregion}

\begin{figure}[bt]
\centering
\includegraphics[width=0.9\linewidth]{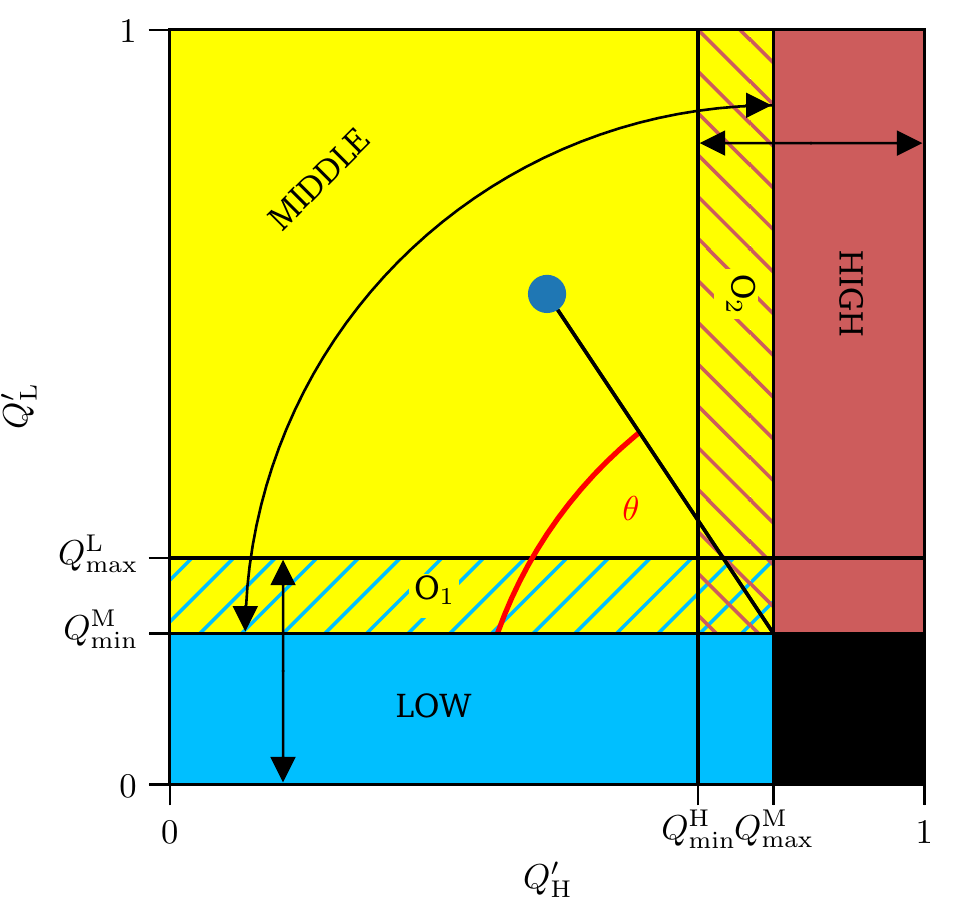}
\caption{Visualization of the different regions: Low (blue), Middle
  (yellow), High (red), overlap $\mathrm{O}_1$ (blue hatched), and
  overlap $\mathrm{O}_2$ (red hatched). A generic configuration in the
  middle region is depicted by a blue point. The arrows indicate the
  ``transitions'' that should be enhanced by reweighting in the
  respective regions. The black region is the region where the system
  is simultaneously both in the high and the low region. This region
  is forbidden and those configurations are strictly rejected.}
\label{fig:regionssketch}
\end{figure}

\begin{figure}[thbp]
  \centering
  \includegraphics[width=0.8\linewidth]{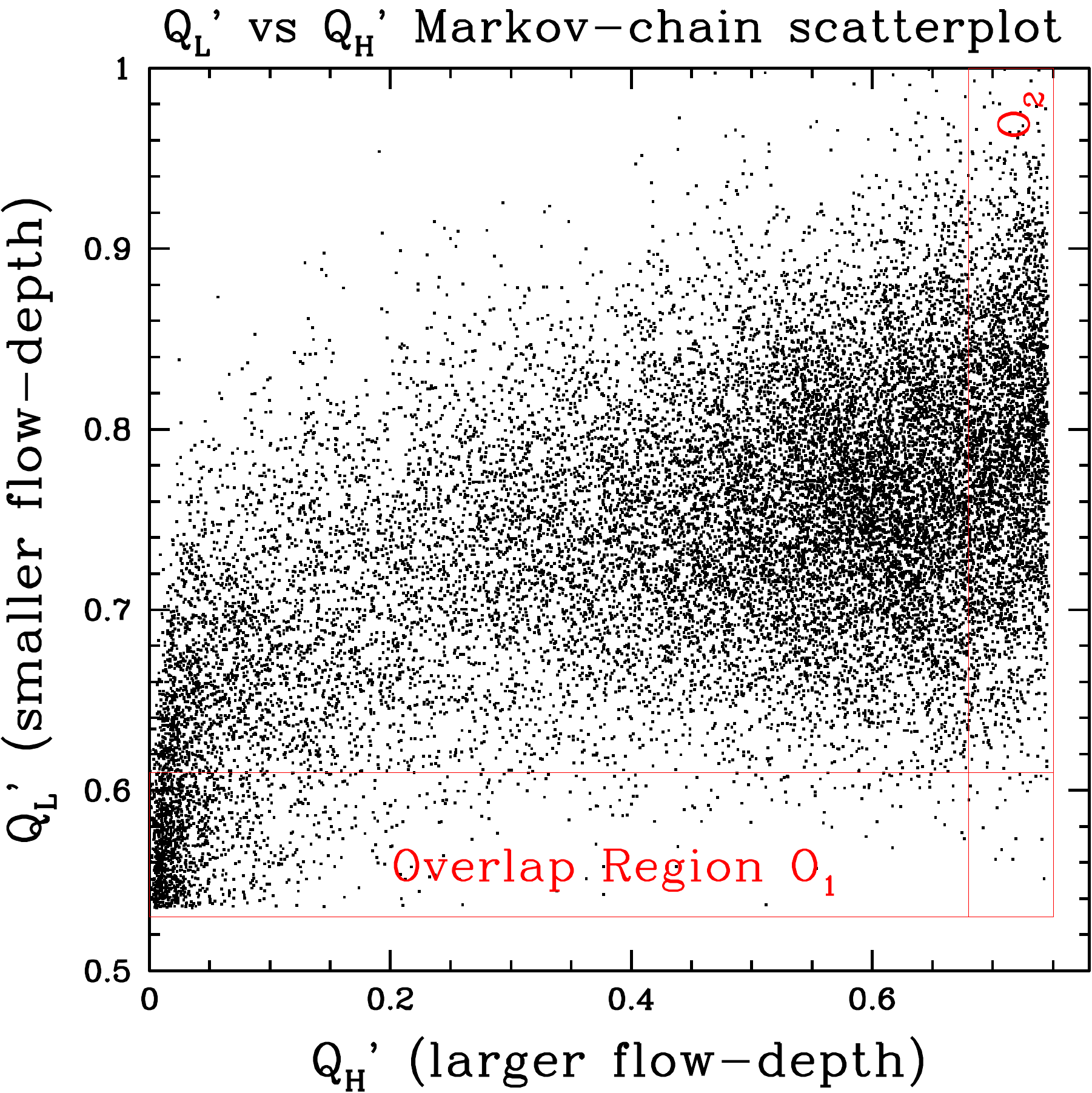}
  \caption{Scatter plot of $\Qpl$ against $\Qph$ in a reweighted
    Markov chain sampling of the middle region, in a $14\times 48^3$
    lattice at $T=7.0 \Tc$.  The values are strongly but
    imperfectly correlated, and very few points land in the overlap
    of $\mathrm{O}_1$ and $\mathrm{O}_2$.}
  \label{fig:Q1Q2}
\end{figure}

The \emph{middle region} (M) is chosen such that it has an overlap
with both the high and the low regions, while those regions are
disjoint. In this region, we measure the topological charge after both
flow times, \textit{i.e.}, we measure both $\Qpl$ and $\Qph$.
These quantities are
highly correlated but still different, and the middle region aims to
smoothly transition from one to the other. This then corresponds to
smoothly connecting the low and high regions. The middle region is
constrained by requiring
$\Qpl > \QMmin \equiv \Qlmax/1.15$ and
$\Qph < \QMmax \equiv 1.15 \times \Qhmin$,
meaning that the overlap with both
the high and low regions is 15\%. Configurations with
$\Qpl < \QMmin$ or $\Qph > \QMmax$
are strictly rejected.  The remaining values are reweighted
according to a reweighting function which smoothly interpolates
between being purely dependent on $\Qpl$ at $\QMmin$ and being purely
dependent on $\Qph$ at $\QMmax$.  Specifically, we define
the reweighting variable $\theta$ to be the angle
\begin{align}
\theta = \arctan\left( \frac{\Qpl - \QMmin}{\QMmax - \Qph} \right) \in
\left[ 0,\, \pi/2 \right] \,. 
\end{align}
$\theta=0$ then corresponds to $\Qpl=\QMmin$ which connects the middle
region with the low region, and $\theta = \nicefrac \pi 2$ corresponds
to $\Qph = \QMmax$ which connects the middle region with the high
region. The overlap regions are then defined as
\begin{align}
  \mathrm O_1 &\equiv \left\{
  \Qpl \colon \QMmin \le \Qpl \le \Qlmax \right\} \,,
\\
\mathrm O_2 &\equiv \left\{
  \Qph \colon \Qhmin \le \Qph \le \QMmax \right\} \,.
\end{align}
The different regions are visualized in
Fig.~\ref{fig:regionssketch}. In the middle region, using an HMC
trajectory of four steps of length $0.25 \, a$ turned out to give the
best performance.  Note that $\Qpl$ and $\Qph$ are highly correlated,
so few if any configurations lie in the lower right part of the
figure; in fact, our method implicitly assumes that the black region
is empty.  Figure \ref{fig:Q1Q2} shows a scatterplot of $\Qpl$ and
$\Qph$ for a reweighted Markov chain performed in this middle region,
showing that the two $Q'$ values are correlated, and that almost no
configurations land in the overlap of the two overlap regions.

The statistical power of a Monte-Carlo in this middle region improves
quite quickly with the length of the Markov chain, because the
reweighting proves to be modest.  Therefore one can build a sample
with negligible statistical errors using a fraction of the Monte-Carlo
time needed on the other regions.

\subsection{Reweighting with Multiple Regions}

Our strategy will be to perform one independent Monte-Carlo simulation
in each of the three regions.  In each Monte-Carlo, we record the
reweighting function $W$ and the true topology $Q$ for every tenth
configuration generated by the Markov chain.  Here we show how these
independent Monte-Carlos can be combined to determine the topological
susceptibility.

We have a sample of $Q',W[Q'],Q$ values from each Markov chain, and we
want to use them to determine the susceptibility.  As we see in
Eq.~(\ref{eq:chidef}), we need to determine $\langle Q^2 \rangle$;
because $|Q| \geq 2$ configurations are negligible, this is the same
as the fraction of configurations with $Q=1$.  In a single-region
Monte Carlo simulation, we would determine that via
\begin{align}
  \label{FQ}
\left\langle Q^2 \right\rangle &\equiv \frac{\int \mathcal{D}U \:\e^{-\beta S[U]} \; \Theta\left(Q - Q_\mathrm{thresh}\right)}
{\int \mathcal{D}U \: \e^{-\beta S[U]}} \nonumber
\\
&\simeq \frac{\sum_i \e^{-W[Q'_i]} \: \Theta\left(Q_i-Q_\mathrm{thresh}\right)}{\sum_i \e^{-W[Q'_i]}} \,,
\end{align}
where $Q'_i$ is the determined $Q'$ value for the $i$ configuration in
the sample.  Here $Q_i$ is $|Q|$ measured on the $i$'th configuration
after some (larger) depth of gradient flow, and $Q_{\mathrm{thresh}}$
is a threshold used to separate the $Q=1$ and $Q=0$ sectors.  We will
check later that the specific values of flow depth and threshold have
almost no bearing at the lattice spacings we consider.

This approach now has to be extended for multiple regions
with their own Monte Carlo samples. The key is the correct use of the
overlap regions. We introduce the shorthand notation $P_R$ for the
fraction of the total probability over all configurations, which lies
in region $R$. That is,
\begin{align}
\label{PR}
P_R = \frac{\int \mathcal{D} U \:\e^{-\beta S[U]} \; \Theta[Q' \in R]} {\int \mathcal{D} U \: \e^{-\beta S[U]}} \,,
\end{align}
where $\Theta[Q' \in R]$ means that we include only those configurations which satisfy the condition to be in region $R$. Similarly, we introduce $P_{R,\,Q}$ to mean the same but with the additional requirement that $Q=1$:
\begin{align}
\label{PRQ}
P_{R,\,Q} = \frac{\int \mathcal{D} U \: \e^{-\beta S[U]} \; \Theta[Q' \in R] \; \Theta\left(Q - Q_\mathrm{thresh}\right)}{\int \mathcal{D} U \: \e^{-\beta S[U]}} \,.
\end{align}
Defining A (``All'') to be the region containing the whole reweighting domain, we need to determine
\begin{align}
\label{FQ1}
\left\langle Q^2 \right\rangle = \frac{P_{\mathrm A,\,Q}}{P_\mathrm A} = \frac{P_{\mathrm L,\,Q} + P_{\mathrm M-\mathrm O_1-\mathrm O_2,\,Q} + P_{\mathrm H,\,Q} }{P_\mathrm L+P_{\mathrm M-\mathrm O_1-\mathrm O_2}+P_\mathrm H}\,,
\end{align}
where we used that $\mathrm A=\mathrm L\cup \left( \mathrm M-\mathrm
O_1-\mathrm O_2 \right) \cup \mathrm H$ and by $\tilde M \equiv
\mathrm M-\mathrm O_1-\mathrm O_2$ we mean all points in the middle
region with both overlap regions removed. Note that this corresponds
to removing the region $\mathrm O_1 \cap \mathrm O_2$ twice,
\textit{i.e.}, $\tilde M$ contains all points in the middle region
that are in neither of the two overlap regions with weight $+1$, the
disjoint parts of the two overlap regions with weight $0$, and the
common part of the overlap regions, \textit{i.e.}, $\mathrm O_1 \cap
\mathrm O_2$, with weight $-1$. Eq.~\eqref{FQ1} can be rewritten as
\begin{align}
\left\langle Q^2 \right\rangle &= \frac{P_{\mathrm H,\,Q}}{P_\mathrm L} \times \left(\frac{P_{\mathrm L,\,Q}+P_{\mathrm M-\mathrm O_1-\mathrm O_2,\,Q}+P_{\mathrm H,\,Q}}{P_{\mathrm H,\,Q}} \right) \nonumber
\\
&\phantom{=\frac{P_{\mathrm H,\,Q}}{P_\mathrm L}} \ \times \left( \frac{P_\mathrm L + P_{\mathrm M-\mathrm O_1-\mathrm O_2} + P_\mathrm H}{P_\mathrm L} \right)^{-1} \nonumber
\\
&\equiv \Xi_0 \cdot \Xi_1 \cdot \Xi_2^{-1} \,.
\end{align}
Using the overlap regions, each of those terms can be rewritten as
\begin{subequations}
\begin{align}
\Xi_0 &\equiv \frac{P_{\mathrm H,\,Q}}{P_\mathrm L} = \frac{P_{\mathrm H,\,Q}}{P_{\mathrm O_2}} \times \frac{P_{\mathrm O_2}}{P_{\mathrm O_1}} \times \frac{P_{\mathrm O_1}}{P_\mathrm L} \,,
\\
\Xi_1 &\equiv \frac{P_{\mathrm L,\,Q}+P_{\mathrm M-\mathrm O_1-\mathrm O_2,\,Q}+P_{\mathrm H,\,Q}}{P_{\mathrm H,\,Q}} \nonumber
\\
\begin{split}
&= 1 + \left( \frac{P_{\mathrm M-\mathrm O_1-\mathrm O_2,\,Q}}{P_{\mathrm O_2}} \times \frac{P_{\mathrm O_2}}{P_{\mathrm H,\,Q}} \right)
\\
&\phantom{= 1} \ + \left( \frac{P_{\mathrm L,\,Q}}{P_{\mathrm O_1}} \times \frac{P_{\mathrm O_1}}{P_{\mathrm O_2}} \times \frac{P_{\mathrm O_2}}{P_{\mathrm H,\,Q}}\right) \,,
\end{split}
\\
\Xi_2 &\equiv \frac{P_\mathrm L + P_{\mathrm M-\mathrm O_1-\mathrm O_2} + P_\mathrm H}{P_\mathrm L} \nonumber
\\
\begin{split}
&= 1 + \left( \frac{P_{\mathrm M-\mathrm O_1-\mathrm O_2}}{P_{\mathrm O_1}} \times \frac{P_{\mathrm O_1}}{P_\mathrm L} \right)
\\
&\phantom{= 1} \ + \left( \frac{P_\mathrm H}{P_{\mathrm O_2}} \times \frac{P_{\mathrm O_2}}{P_{\mathrm O_1}} \times \frac{P_{\mathrm O_1}}{P_\mathrm L} \right) \,,
\end{split}
\end{align}
\end{subequations}
where now each ratio is determined by a single Monte Carlo simulation in the high region, middle region, or low region.

We expect that almost all of the total weight of configurations is in
the low region, while almost all of the total weight of $Q=1$
configurations lies in the high region. Consequently, we expect that
$\Xi_1 \approx 1 \approx \Xi_2$.  Naturally we shall check this; but
to the extent that it holds, we obtain the easier expression
\begin{align}
\left\langle Q^2 \right\rangle \approx \Xi_0 = \frac{P_{\mathrm
    H,\,Q}}{P_{\mathrm O_2}} \times \frac{P_{\mathrm O_2}}{P_{\mathrm
    O_1}} \times \frac{P_{\mathrm O_1}}{P_\mathrm L} \,.
\label{eq:qsquaredapprox_glob}
\end{align}
Under this approximation, we end up with a product of three ratios,
each of which can be determined using a single one of our Monte Carlo
samples:
\begin{align}
\frac{P_{\mathrm H,\,Q}}{P_{\mathrm O_2}} &= \frac{\sum_{i\in \mathrm
    H} \: \e^{-W_\mathrm{high}\left[\Qph\right]} \;
  \Theta\left(Q-Q_\mathrm{thresh} \right)} {\sum_{i\in \mathrm H} \:
  \e^{-W_\mathrm{high}\left[\Qph\right]} \;
  \Theta\left(\Qph \in \mathrm O_2\right)} \,,
%\\\nonumber
\\
\frac{P_{\mathrm O_2}}{P_{\mathrm O_1}} &= \frac{\sum_{i\in \mathrm M}
  \: \e^{-W_\mathrm{mid}\left[\theta\right]} \;
  \Theta\left(\Qph \in \mathrm O_2\right)} {\sum_{i\in \mathrm M} \:
  \e^{-W_\mathrm{mid}\left[\theta\right]} \; \Theta\left(\Qpl
  \in \mathrm O_1\right)} \,,
%\\\nonumber
\\
\frac{P_{\mathrm O_1}}{P_\mathrm L} &= \frac{\sum_{i\in \mathrm L} \:
  \e^{-W_\mathrm{low}\left[\Qpl\,,G\right]} \; \Theta(\Qpl \in
  \mathrm O_1)} {\sum_{i\in \mathrm L} \:
  \e^{-W_\mathrm{low}\left[\Qpl\,,G\right]}} \,.
\end{align}
Therefore, within our approximations, we can easily determine all
three ratios. And we need all three Monte Carlos, because each
determines one of these three ratios.

\subsection{Parameters to Tune}

As in the original reweighting approach, we still find that a certain
amount of hand-tuning is required to achieve the best efficiency of
our method. First, there are the depths of gradient flow to use in
establishing the reweighting variables $\Qpl$, $G$, and $\Qph$. We
find that in the high region a rather large amount of gradient flow is
required to carefully distinguish calorons of different sizes. In the
low region, we need to tune both $t_{f,\mathrm{low}}$ and $t_{f,G}$, and it is
not clear that both flow depths should be the same. In particular,
$t_{f,G}$ needs to be enough flow that fluctuations are removed, but too
much flow shrinks the dislocations and the peak action density cannot
distinguish between trivial topology and dislocations any
more. Similar arguments hold for $t_{f,\mathrm{low}}$. We find that
$t_{f,\mathrm{low}}$ slightly larger than $t_{f,G}$ improves the efficiency,
but a more careful analysis would be desirable, especially in view of
the inclusion of fermions. This could be done by carefully comparing
the peak action density of discretized calorons and thermal
configurations after different amounts of gradient flow.

Second, there is the position of the lower bound of the high
region. We chose $\Qhmin=0.7$ throughout because
this choice definitely includes the high barrier for all
lattices we consider. Changing $\Qhmin$ affects the sampled regions
for both the middle and high analyses, and will impact the statistical
power of each Monte-Carlo in opposite directions.  It might be worth
revisiting what value is optimal overall.

Next, there is the length of the HMC trajectories used in the
respective regions. We chose the lengths such the the acceptance rate
of the reweighting Metropolis step is about 50\%; this leads to small
trajectories in the low region, large trajectories in the high region,
and to intermediate-sized trajectories in the middle region. Again, a
more careful analysis could still improve the efficiency of the
algorithm by comparing the achieved statistics at fixed numerical
effort as a function of the HMC trajectory lengths in the respective
regions.

Finally, there is the definition of the topological charge as the
observable for determining the topological susceptibility. Since we
saw in Ref.~\cite{Jahn:2018dke} that the continuum extrapolated
results are insensitive to the exact choices of both the flow depth
and the threshold, we use $t_f=2.4 \, a^2$ of Wilson flow and
$Q_\mathrm{thresh} = 0.7$ for deciding whether a configuration is
topological or not throughout this section, in accordance with the
choices in Ref.~\cite{Jahn:2018dke}. This allows us to directly
compare our results to the ones obtained with the original reweighting
approach.  We will investigate other choices to ensure that our final
answers are not dependent on this choice.

\section{Results \label{sec:globresults}}

\begin{table*}[t]
\centering
\caption{The lattices used in this section. The lattices labeled with
  ``A'' correspond to simulations at $2.5\,\Tc$, the lattices labeled
  with ``B'' correspond to $4.1\,\Tc$, and the lattices labeled with
  ``C'' are simulations at $7 \,\Tc$.  We also indicate the amount of
  gradient flow that was used in the definitions of each reweighting
  variable.}
\label{tab:ourlattsglob}
%{\textwidth}{c @{\extracolsep{\fill}}
\begin{tabular*}{.85\textwidth}{l @{\extracolsep{\fill}} ccccccc}\hline\hline
Lat 					& 	$T/\Tc$	&
$\phantom{a}N_\tau\phantom{a}$	&	$N_x\times N_y \times N_z$			&	$\beta_\mathrm{lat}$		&	$t_{f,\mathrm{low}}/a^2$	&	$t_{f,G}/a^2$	&	$t_{f,\mathrm{high}}/a^2$		\\\hline

$\text{A}_1$			&	2.5		&	10							&	$36 \times 32^2$	&	$6.90097$			&	$0.48$			&	$0.42$		&	$0.96$				\\
$\text{A}_2$			&	2.5		&	12							&	$40 \times 36^2$	&	$7.04966$			&	$0.48$			&	$0.48$		&	$0.96$				\\
$\text{A}_3$			&	2.5		&	14							&	$48^3$			&	$7.17706$			&	$0.54$			&	$0.48$		&	$1.32$				\\\hline

$\text{B}_1$			&	4.1		&	10							&	$36 \times 32^2$	&	$7.30916$			&	$0.42$			&	$0.42$		&	$0.96$				\\
$\text{B}_2$			&	4.1		&	12							&	$40 \times 36^2$	&	$7.46275$			&	$0.48$			&	$0.48$		&	$0.96$				\\
$\text{B}_3$			&	4.1		&	14							&	$48^3$			&	$7.59354$			&	$0.48$			&	$0.48$		&	$1.32$				\\\hline

$\text{C}_{1\mathrm a}$	&	7.0		& 	10							&	$12^3$			&	$7.76294$			&	$0.36$			&	$0.36$		&	$0.96$				\\
$\text{C}_{1\mathrm b}$	&	7.0		&	10							&	$16^3$			&	$7.76294$			&	$0.36$			&	$0.36$		&	$0.96$				\\
$\text{C}_{1\mathrm c}$	&	7.0		&	10							&	$24^3$			&	$7.76294$			&	$0.42$			&	$0.42$		&	$0.96$				\\
$\text{C}_{1\mathrm d}$	&	7.0		&	10							&	$32^3$			&	$7.76294$			&	$0.42$			&	$0.42$		&	$0.96$				\\
$\text{C}_{1\mathrm e}$	&	7.0		&	10							&	$40^3$			&	$7.76294$			&	$0.48$			&	$0.48$		&	$0.96$				\\\cmidrule{3-8}
$\text{C}_2$			&	7.0		&	12							&	$40 \times 36^2$	&	$7.91939$			&	$0.48$			&	$0.48$		&	$0.96$				\\
$\text{C}_3$			&	7.0		&	14							&	$48^3$			&	$8.05216$			&	$0.48$			&	$0.48$		&	$1.32$				\\\hline\hline
\end{tabular*}
\end{table*}

Our goal is to demonstrate that the improved reweighting method as
described above yields statistically powerful results in a range of
lattice spacings and volumes and allows for the determination of the
topological susceptibility up to $7.0 \, \Tc$ in the quenched
approximation, where the original reweighting approach from
Ref.~\cite{Jahn:2018dke} is limited due to the barriers described
above. To crosscheck our results, we also determine the susceptibility
again at $2.5 \, \Tc$ and $4.1 \, \Tc$. We already saw in the original
approach that at such high temperatures it is sufficient to only take
into account the $Q=1$ sector because the higher topological sectors
are too suppressed  to significantly contribute to the topological
susceptibility. We therefore only reweight the $Q=1$ sector as
discussed in the previous section.

\begin{figure}[tp]
\centering
\includegraphics[width=\linewidth]{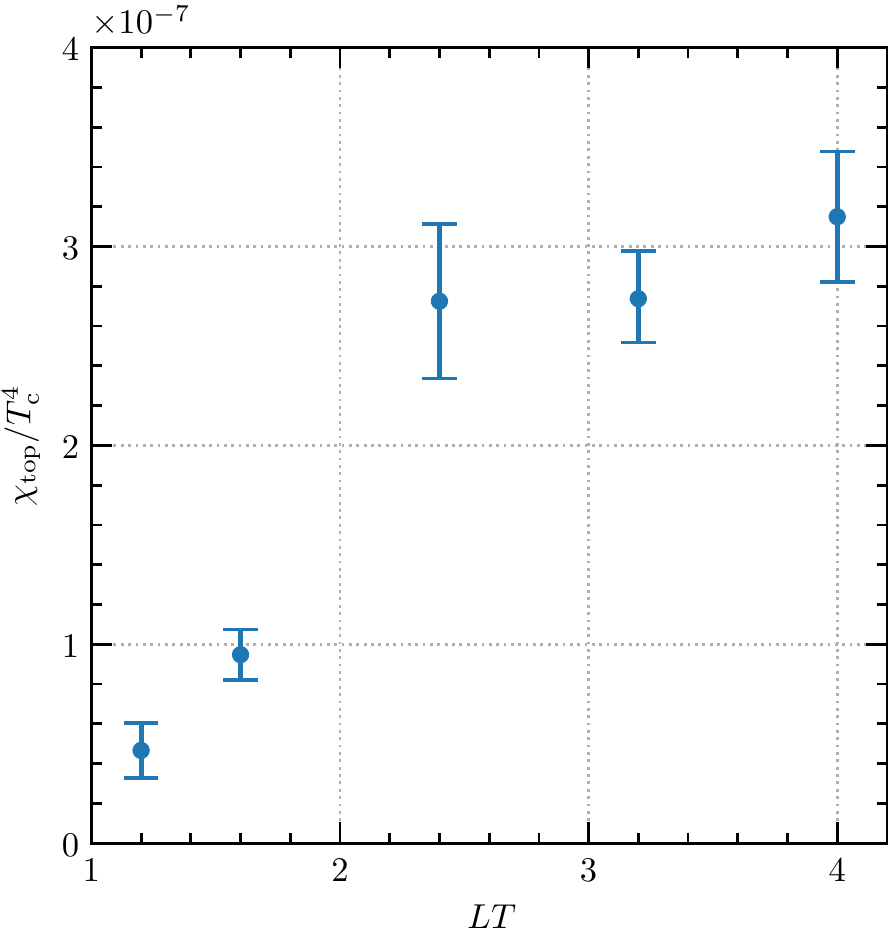}
\caption{Finite volume dependence of the topological susceptibility
  $\chitop$ at $N_\tau=10$ and $T = 7.0\,\Tc$ (Lattices
  C$_{\mathrm{1a}}$ through C$_{\mathrm{1e}}$).}
\label{fig:finite_volume_glob}
\end{figure}

First we investigate the dependence of the susceptibility on the
lattice aspect ratio.  It is known \cite{Braaten:1995cm} that the
long-distance correlations of Yang-Mills theory well above $\Tc$ are
described by a 3D theory with correlation lengths which are
parametrically of order $1/gT$ and $1/g^2 T$.  This implies that, for
all temperatures $T \gg \Tc$, an aspect ratio which is parametrically
$\mathcal{O}(1/g^2)$ should be sufficient; there is no need to keep
the physical volume fixed as we increase the temperature.
Since $1/g^2$ is largest for the highest temperature, we then study
the aspect ratio dependence at $T=7\Tc$ and we assume that a ratio
which is sufficient at this temperature will also work at the lower
temperatures.  We show the resulting susceptibility as a function of
aspect ratio, all at $N_\tau = 10$, in Figure
\ref{fig:finite_volume_glob}.  The results indicate that aspect ratios
above about 2.4 show no discernible volume dependence.  Therefore we
conservatively choose aspect ratios somewhat above 3 in all other
cases.

To carry out the continuum extrapolation we will consider lattice
spacings with $N_\tau = 10,\,12,\,14$ at each temperature we explore.
We adopt the scale-setting (the relation between the lattice spacing
$a$ and inverse coupling $\beta_{\mathrm{latt}}$) determined in
\cite{Burnier:2017bod}.  In total, we study 13 different lattice
setups as listed in Tab.~\ref{tab:ourlattsglob}.  All calculations
were conducted over a three month period on the Lichtenberg high
performance computer center of the TU Darmstadt and on one server node
with four 16-core Xeon Gold CPUs.  Because some machines had 24-core
nodes, we made some unusual choices of lattice sizes such that at
least one lattice direction would be a multiple of 12.

\begin{figure}[t]
\centering
\includegraphics[width=\linewidth]{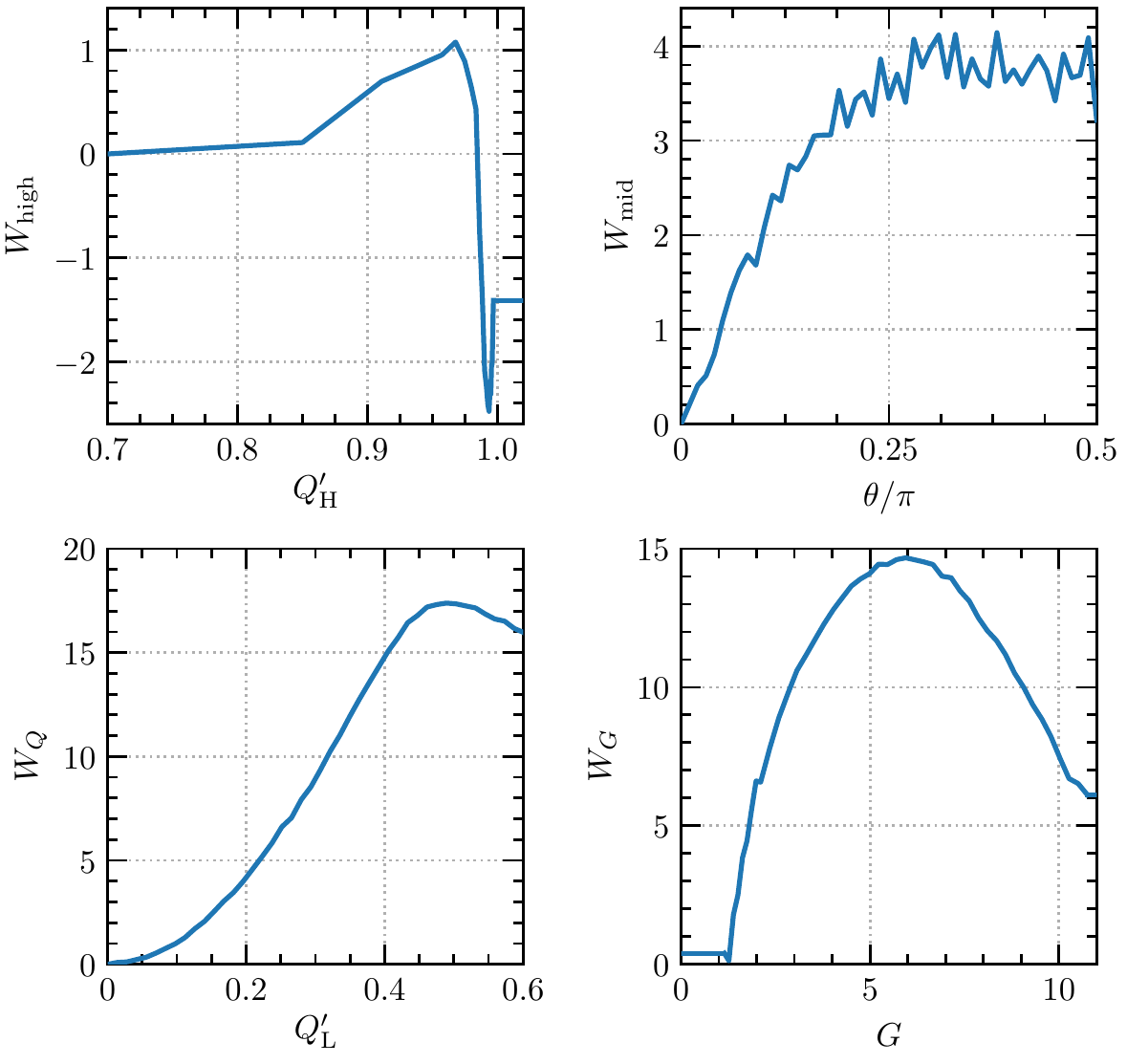}
\caption{Reweighting functions for a $14 \times 48^3$ lattice at $7 \,
  \Tc$ (Lattice C$_3$). The reweighting function in the low region is
  the sum of the two reweighting functions in the low panels,
  \textit{i.e.}, $W_\mathrm{low}(\Qpl,G) = W_Q(\Qpl) + W_G(G)$.}
\label{fig:reweighting_functions_glob}
\end{figure}

The first task is to build the four reweighting functions that are
needed to completely sample one of the lattices. In total, we
therefore have 52 different reweighting functions. One example of
these functions for a $14\times48^3$ lattice at $7 \, \Tc$ is shown in
Fig.~\ref{fig:reweighting_functions_glob}. The reweighting function
$W_\mathrm{high}$ looks, as expected, like the high-$Q'$ part of the
reweighting function in the original reweighting approach. It shows a
very narrow minimum around $\Qph = 1$, corresponding to
genuine $Q = 1$ calorons. At smaller $\Qph$, the reweighting
function shows a plateau corresponding to dislocations. Moving from a
genuine caloron to a dislocation requires a reweighting of only about
$\e^{-3}$ which is the reason that at lower temperatures and coarser
lattices, where the barrier gets even smaller, no reweighting is
needed at all to efficiently sample the high region. The reweighting
function $W_\mathrm{mid}$ samples between dislocations of different
sizes and the function shows a monotonically increasing trend. The
``spikes'' result from the fact that we discretized the $\theta$
domain with 50 intervals; a smaller number would have been
sufficient. However, the reweighting needed to sample this region is
only $\e^{-4}$ and due to the ``simple'' monotonically increasing form
of the reweighting function and the small structural differences
between the relevant configurations, the middle region is sampled very
efficiently. The low region is sampled with the sum of the two
reweighting functions $W_Q$ and $W_G$. The topological-charge
reweighting function $W_Q$ has a deep minimum at $\Qpl=0$,
corresponding to ordinary, topologically trivial $Q=0$
configurations. The function then increases until a maximum is reached
that corresponds to dislocations. Note that reaching the dislocations
requires a large amount%
\footnote{%
  The total reweighting between $Q=0$ configurations and dislocations
  is the sum of this $e^{18}$ reweighting and $e^6$ of reweighting
  between the smallest and largest $G$ values.  This large $e^{24}$
  factor is the reason that simulations without any reweighting are
  fruitless at this temperature.}
of reweighting of roughly $\e^{-18}$. The peak
action-density reweighting function $W_G$ shows a large maximum at
intermediate $G$ which corresponds to the ``gap'' in the right panel
of Fig.~\ref{fig:globbiness}. Moving through this gap therefore
requires a large amount of reweighting of roughly $\e^{-15}$.

With these reweighting functions at hand, we proceed to determine the
topological susceptibility via Eqs.~\eqref{eq:chidef} and
\eqref{eq:qsquaredapprox_glob}. Since we saw in the original
reweighting approach that it does not affect the results if we use
Wilson or Zeuthen flow, we only use the computationally slightly
cheaper Wilson flow here. For deciding whether a configuration is
topological or not, we measure the topological charge after
$t_f = 2.4 \, a^2$ Wilson flow, thresholded with
$Q_\mathrm{thresh}=0.7$.  We will present a check on this procedure at
the end of this section.

The first question is whether the approximations proposed in
Eq.~\eqref{eq:qsquaredapprox_glob} are justified. For this, we tested
the validity of our approximations by explicitly measuring $\Xi_1$ and
$\Xi_2$ on lattices $\mathrm A_\mathrm 1$ and
$\mathrm C_\mathrm{2}$. The result is
\begin{align}
\left| 1 - \Xi_1^{\mathrm A_1} \right| \equiv 0 \equiv \left| 1 - \Xi_1^{\mathrm C_2} \right|
\end{align}
meaning that there is not a single caloron in the middle and low regions. Also the second approximation is fulfilled very precisely:
\begin{align}
\left| 1 - \left( \Xi_2^{\mathrm A_1} \right)^{-1} \right| &\approx 3.2 \times 10^{-4} \,,
\\
\left| 1 - \left( \Xi_2^{\mathrm C_2} \right)^{-1} \right| &\approx 3.6 \times 10^{-9} \,.
\end{align}
That is, all but a tiny fraction of the total weight in the ensemble
is contained in configurations in the low region.
We therefore proceed using Eq.~\eqref{eq:qsquaredapprox_glob} for
determining the topological susceptibility. For completeness, we
present all our results in Tab.~\ref{tab:resultsglob}.

\begin{table*}[t]
\centering
\caption{Number of measurements (with 10 HMC trajectories per
  measurement), number of times the Markov chain moved from the top to
  the bottom of each range and back, and the final determined
  susceptibility for each lattice listed in Table \ref{tab:ourlattsglob}.}
\label{tab:resultsglob}
%{\textwidth}{c @{\extracolsep{\fill}}
\begin{tabular*}{.85\textwidth}{l @{\extracolsep{\fill}} rrrrrrr}\hline\hline
\multirow{2}{*}{Lat} 		&\multicolumn{3}{c}{\#Measurements}								&\multicolumn{3}{c}{\#Complete sweeps}								&\multirow{2}{*}{$\ln\left( \chitop/\Tc^4 \right)$}	\\
					& \multicolumn{1}{c}{L}	&\multicolumn{1}{c}{M} 	& \multicolumn{1}{c}{H} 	&\multicolumn{1}{c}{L}	& \multicolumn{1}{c}{M}	&\multicolumn{1}{c}{H} 										\\\hline

$\text{A}_1$			&	$175{,}560$		&	$40{,}680$		&	$53{,}820$		&	536				&	861				&	805				&	$-8.11(09)$						\\
$\text{A}_2$			&	$139{,}690$		&	$16{,}670$		&	$62{,}980$		&	464				&	491				&	554				&	$-8.23(12)$						\\
$\text{A}_3$			&	$182{,}460$		&	$24{,}790$		&	$56{,}760$		&	665				&	626				&	592				&	$-8.47(10)$						\\\hline

$\text{B}_1$			&	$237{,}960$		&	$32{,}440$		&	$61{,}120$		&	768				&	593				&	970				&	$-11.41(09)$						\\
$\text{B}_2$			&	$337{,}540$		&	$16{,}940$		&	$71{,}530$		&	479				&	453				&	625				&	$-11.63(11)$						\\
$\text{B}_3$			&	$167{,}990$		&	$21{,}410$		&	$59{,}030$		&	490				&	549				&	526				&	$-11.83(12)$						\\\hline

$\text{C}_{1\mathrm a}$	&	$100{,}000$		&	$100{,}000$		&	$66{,}400$		&	83				&	$2{,}215$			&	$1{,}915$			&	$-16.88(30)$						\\
$\text{C}_{1\mathrm b}$	&	$100{,}000$		&	$100{,}000$		&	$74{,}200$		&	105				&	$2{,}450$			&	$1{,}822$			&	$-16.17(13)$						\\
$\text{C}_{1\mathrm c}$	&	$321{,}700$		&	$26{,}300$		&	$24{,}200$		&	269				&	454				&	385				&	$-15.12(14)$						\\
$\text{C}_{1\mathrm d}$	&	$368{,}420$		&	$21{,}630$		&	$84{,}800$		&	483				&	655				&	$1{,}353$			&	$-15.11(08)$						\\
$\text{C}_{1\mathrm e}$	&	$307{,}300$		&	$15{,}800$		&	$41{,}700$		&	422				&	382				&	636				&	$-14.97(10)$						\\\cmidrule{2-8}
$\text{C}_2$			&	$487{,}240$		&	$32{,}080$		&	$54{,}420$		&	490				&	931				&	496				&	$-15.22(11)$						\\
$\text{C}_3$			&	$268{,}220$		&	$20{,}360$		&	$60{,}920$		&	462				&	475				&	531				&	$-15.45(12)$						\\\hline\hline
\end{tabular*}
\end{table*}

\begin{figure*}
\centering
\includegraphics[width=\linewidth]{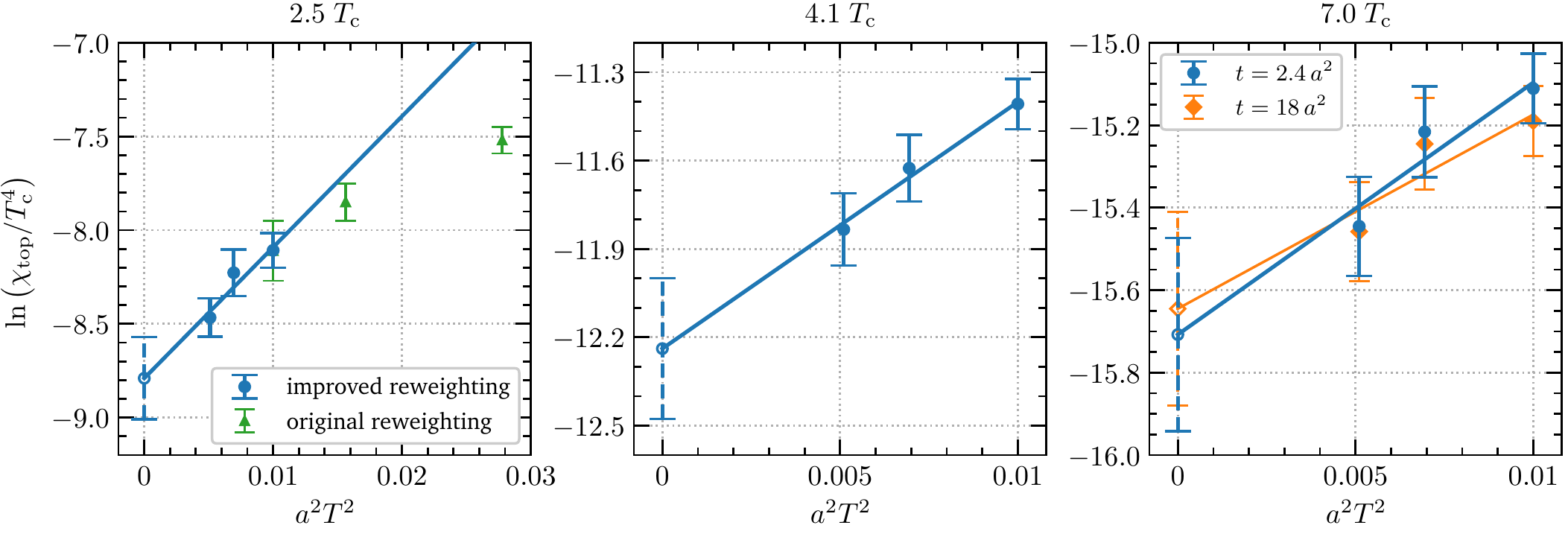}
\caption{Continuum extrapolation in terms of the logarithm of the
  topological susceptibility $\chitop$ for the three
  temperatures. Left: $2.5 \, \Tc$. We additionally show the results
  from the original reweighting approach \cite{Jahn:2018dke}. Middle:
  $4.1 \, \Tc$. Right: $7 \, \Tc$, using the lattices C$_\mathrm{1d}$,
  C$_2$, and C$_3$. We additionally show the results with flow time
  $t_f=18.0 \, a^2$. The points with dashed error bars are the
  continuum-extrapolated results.}
\label{fig:continuumextrap_glob}
\end{figure*}

We now consider the volume dependence by studying $\chitop$ as a
function of the aspect ratio at $7 \, \Tc$ with $N_\tau = 10$, using
lattices C$_{\mathrm{1a}}$ through C$_{\mathrm{1e}}$. This is
presented in Fig.~\ref{fig:finite_volume_glob}. In agreement with the
corresponding result in the original reweighting approach, aspect
ratios smaller than 2 are badly discrepant, while aspect ratios larger
than about 2.5 give consistent results and the large-volume behavior
is reached. For determining the continuum extrapolation of the
topological susceptibility, we therefore use aspect ratios between 3
and 3.5.

Finally, we address the continuum extrapolation of the topological
susceptibility at three temperatures $2.5 \, \Tc$, $4.1 \, \Tc$, and
$7.0 \, \Tc$, using three lattice spacings for each temperature with
$N_\tau = 10,\,12,\,14$. As already elaborated in the discussion of
the continuum extrapolation of the original reweighting approach, the
continuum extrapolation is conducted in terms of the logarithm of the
susceptibility, \textit{i.e.}, we linearly extrapolate $\ln \left(
\chitop \right)$ against $a^2$. The continuum extrapolations of the
three different temperatures are presented in
Fig.~\ref{fig:continuumextrap_glob}. At $T=2.5 \, \Tc$ (left panel of
Fig.~\ref{fig:continuumextrap_glob}), we additionally show the results
with $t_f=2.4 \, a^2$ Wilson flow from the original reweighting
approach. This clearly shows that lattices with $N_\tau=6$ are too
coarse to be in the scaling region, while the finer lattices with
$N_\tau=8$ and $N_\tau=10$ are consistent with the continuum
extrapolation. To explicitly check that the continuum extrapolated
results do not depend on the depth of gradient flow used for
determining the topological charge, we additional show the results for
a much larger depth of flow, $t_f=18.0 \, a^2$, in the $T=7.0 \, \Tc$
continuum extrapolation (right panel of
Fig.~\ref{fig:continuumextrap_glob}). In accordance with our
findings in Ref.~\cite{Jahn:2018dke}, the different $Q$ definitions
are nearly indistinguishable at finer lattices ($N_\tau = 12,14$),
while the difference becomes larger at coarser lattices
($N_\tau=10$). However, the continuum extrapolated results differ only
by about 6\%, despite the very different flow times.

\section{Discussion \label{sec:globdiscussion}}

We presented an extension of the reweighting technique developed in
Ref.~\cite{Jahn:2018dke} that improves the efficiency in determining
the topological susceptibility in pure SU(3) Yang-Mills theory. The
method is based on the individual treatment of the topologically
trivial sector $Q= 0$ (``low region''), the topological $Q=1$ sector
(``high region''), and the intermediate dislocations with fractional
$0 < Q < 1$ (``middle region''). In the low region, a combined
reweighting in terms of the topological charge and the peak action
density, both evaluated at a small amount of gradient flow, allows to
efficiently sample between topologically trivial configurations and
dislocations. Since this requires a large amount of reweighting and
the involved topological objects are small and fragile, only very
small HMC trajectories (one step of $0.2 \, a$) can be used; otherwise
the acceptance rate in the reweighting step becomes very small. In the
high region, sampling between genuine calorons and dislocations
requires only a small amount of reweighting. Since also the involved
topological objects are large and robust, large HMC trajectories
(eight steps of $0.25 \, a$) allow for an efficient sampling. At
coarser lattices ($N_\tau = 10,\,12$), no reweighting is necessary at
all in this region; at finer lattices ($N_\tau=14$), a large amount
$t'_\mathrm H = 1.32 \, a^2$ of Wilson flow allows for a careful
distinction of calorons of different sizes. In the middle region,
sampling between dislocations with different sizes requires only a
small amount of reweighting and using intermediate-size HMC
trajectories (four steps of $0.25 \, a$) allows for a very efficient
sampling.

This method is very effective and allows for a continuum-extrapolated
determination of the topological susceptibility up to the very high
temperature $T= 7.0 \, \Tc$ and up to an aspect ratio of 4 and fine
lattice spacings with $N_\tau = 14$; this determination constitutes
the first direct measurement of the susceptibility at such a high
temperature. Our final results are
\begin{align}
\begin{split}
\chitop(T=2.5\,\Tc) &= 1.52 \times 10^{-4} \ \e^{\pm 0.22} \, \Tc^4\,,
\\
\chitop(T=4.1\,\Tc) &= 4.84 \times 10^{-6} \ \e^{\pm 0.24} \, \Tc^4 \,,
\\
\chitop(T=7.0\,\Tc) &= 1.51 \times 10^{-7} \ \e^{\pm 0.23} \, \Tc^4 \,.
\end{split}
\end{align}
To emphasize again how important reweighting is to a \textsl{direct}
determination of $\chi(T=7.0\,\Tc)$, note that without reweighting,
$Q=1$ configurations in our finest-spacing, highest-temperature
lattice would represent approximately $3\times 10^{-9}$ of all
configurations.  Because there is also a large autocorrelation time to
jump from $Q=1$ back to $Q=0$ configurations, it would take at least
$10^{10}$ HMC updates to observe \textsl{any} topological
configurations, and $\sim 10^{12}$ HMC updates to gather comparable
statistics to what we achieve with $3\times 10^6$ HMC updates.
Therefore, to obtain the susceptibility at such temperatures, either
our reweighting method or an indirect method such as the approach of
Borsanyi \textsl{et al} \cite{Borsanyi:2016ksw} must be used.

The continuum-extrapolated results for $2.5 \, \Tc$ and $4.1 \, \Tc$
are consistent within errors with the corresponding results from the
original reweighting approach \cite{Jahn:2018dke} and hence also with
the literature \cite{Borsanyi:2015cka,Berkowitz:2015aua}. Applying the
grand continuum fit of Ref.~\cite{Borsanyi:2015cka}, also the
continuum extrapolated result at $7 \, \Tc$ agrees well with their
findings.

\begin{figure}[t]
\centering
\includegraphics[width=\linewidth]{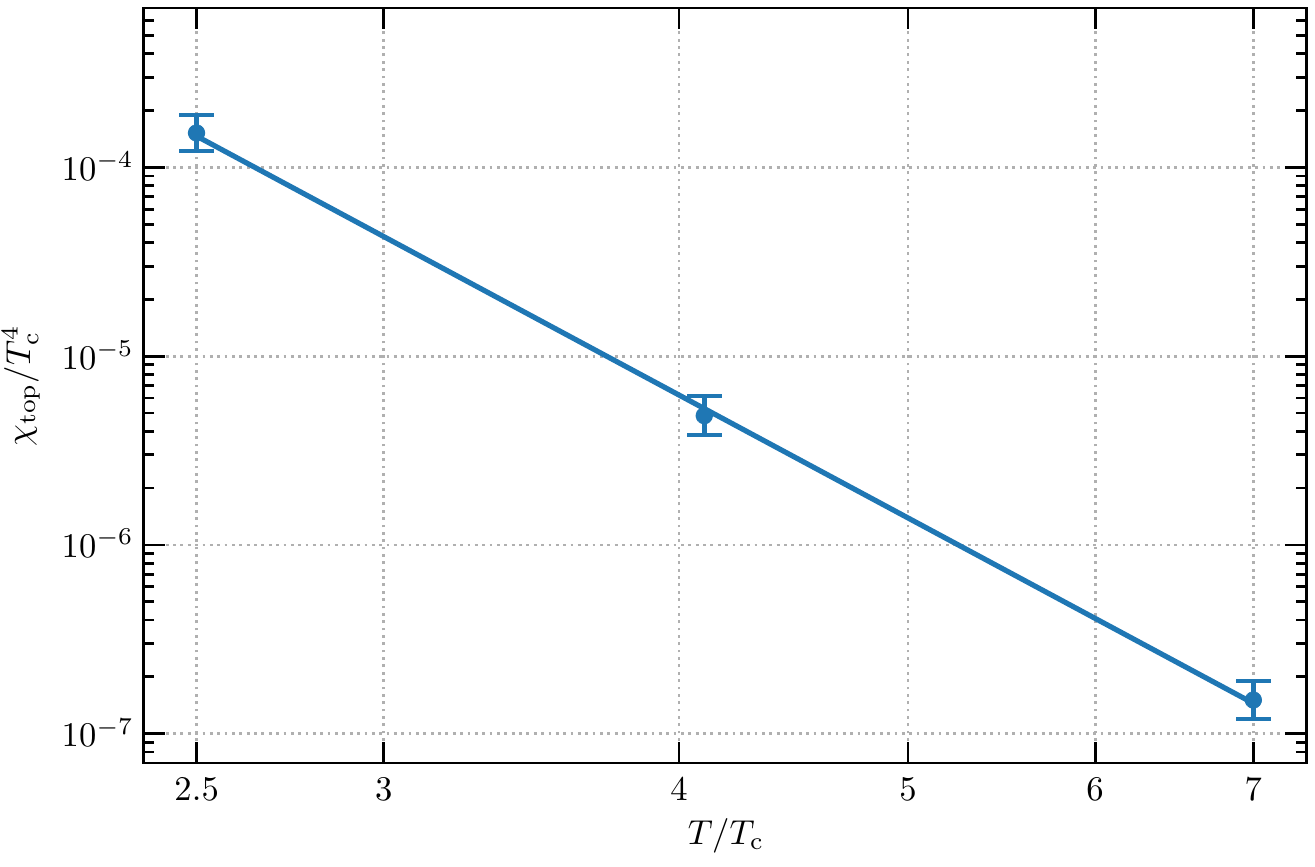}
\caption{Continuum extrapolated results of the topological
  susceptibility as a function of temperature in a double-logarithmic
  plot. We also show the simple power-law fit
  Eqs.~\eqref{eq:powerlawfit} and \eqref{eq:fitparameters}.}
\label{fig:chitop_scaling_glob}
\end{figure}

The continuum extrapolated results are plotted against temperature in
a double-logarithmic plot in Fig.~\ref{fig:chitop_scaling_glob}. A
perturbative prediction of the temperature dependence of the
topological susceptibility at high temperatures in the framework of
the dilute instanton gas approximation \cite{Gross:1980br} suggests
that the continuum extrapolated data may be fitted in the form
\begin{align}
\frac{\chitop}{\Tc^4} = \frac{\chitop^0}{\Tc^4} \left( \frac{T}{\Tc} \right)^b
\label{eq:powerlawfit}
\end{align}
which has a linear behavior in a double logarithmic plot. The best fit parameters of our continuum extrapolated results are
\begin{align}
\ln \left( \frac{\chitop^0}{\Tc^4} \right) = -2.67(46) \,, \qquad b = -6.72(31) \,,
\label{eq:fitparameters}
\end{align}
where the uncertainties are statistical only. In pure SU(3) Yang-Mills
theory, the dilute instanton gas approximation predicts the
exponent $b=-7$ which is consistent with
our result. In a recent study, Borsanyi \emph{et al.}\ also determined
this exponent in a conventional heat-bath/overrelaxation setup
\cite{Borsanyi:2015cka}. Despite applying much more numerical effort,
they have significantly larger statistical errors and only reach $4 \,
\Tc$; but their result $b=-7.1(4)$ is consistent with ours. In their
determination, Berkowitz \emph{et al.}\ found the result $b=-5.64(4)$
\cite{Berkowitz:2015aua} which differs significantly from our
result. However, they only reached $2.5 \, \Tc$, which may be too low
to display the same slope as in the regime we study.

This methodology can be applied in a straightforward way to the full
(unquenched) theory.  However, the $Q=0$ and $Q=1$ sectors differ in
that the former will have no small fermionic eigenvalues, while the
latter will; this may require a careful handling of the fermionic
mass, and the use of fermionic implementations with excellent
chirality properties.  We leave the application of our methodology to
the full theory for future work.

\begin{acknowledgments}
The authors acknowledge support by the Deutsche Forschungsgemeinschaft
(DFG, German Research Foundation) through the CRC-TR 211
``Strong-interaction matter under extreme conditions'' -- project
number 315477589 -- TRR 211. We also thank the GSI Helmholtzzentrum
and the TU Darmstadt and its Institut f\"ur Kernphysik for supporting
this research. Calculations were conducted on the Lichtenberg
high performance computer of the TU Darmstadt. This work was performed
using the framework of the publicly available openQCD-1.6 package
\cite{openQCD}.
\end{acknowledgments}

\bibliographystyle{apsrev4-1}
\bibliography{refs}

\end{document}